# 基于开源 Matrix 指令集扩展（矢量点积）的高性能 RISC-V 处理器"香山"（nanhu 版本）的 LLM 加速的研究[*]

陈煦豪[1,2,4], 胡思鹏[1], 刘洪超[1,2,3], 刘伯然[4], 唐丹[1], 赵地[4,5]

[1](北京开源芯片研究院,北京　100080)
[2](上海科技大学 信息学院,上海　210210)
[3](郑州大学 河南先进技术研究院,郑州 河南　450003)
[4](中国科学院计算技术研究所 处理器芯片全国重点实验室,北京　100190)
[5](中国科学院大学, 北京　100049)
通讯作者：唐丹, E-mail: tangdan@bosc.ac.cn

**摘　要**：　鉴于边缘 AI 的高性能与低功耗需求，本课题基于 RISC-V 指令集架构，针对边缘设备数字信号处理的实际问题，设计了一种边缘 AI 的专用指令集处理器。本设计在有限的硬件开销下，提升了边缘 AI 的执行效率，降低了边缘 AI 的能量消耗，能够满足边缘 AI 应用中进行高效大语言模型 (LLM) 推理计算的需求。本文的主要工作如下：1) 针对大语言模型的特性，基于 RISC-V 指令集扩展了自定义指令完成矢量点积计算，在专用的矢量点积加速硬件上进行大语言模型的运算加速；2) 基于开源高性能 RISC-V 处理器核"香山"nanhu 版本架构，实现了矢量点积专用指令集处理器 nanhu-vdot，它在高性能处理器"香山"(nanhu 版本)的基础上增加了矢量点积计算单元以及流水线处理逻辑；3) 　对 nanhu-vdot 进行 FPGA 硬件测试，在几乎没有硬件资源和功耗的额外消耗条件下，矢量点积运算速度对比标量方法提高 4 倍以上，使用软硬件协同方案进行第二代生成式预训练（GPT-2，Generative Pre-Trained-2）模型推理，相比纯软件实现，速度提高了约 30%。

**关键词**：　指令集扩展;矢量点积;软硬件协同；大语言模型推理

## Research on LLM Acceleration Using the High-Performance RISC-V Processor "Xiangshan" (Nanhu Version) Based on the Open-Source Matrix Instruction Set Extension (Vector Dot Product)

CHEN Xu-Hao[1,2,4],　HU Si-Peng[1],　LIU Hong-Chao[1,3,4],　LIU Bo-Ran[4],　TANG Dan[1],　ZHAO Di[4,5]

[1](Beijing Institute of Open Source Chip, Beijing 100080, China)
[2](School of Information Science and Technology, ShanghaiTech University, Shanghai 210210, China)
[3](Henan Institute of Advanced Technology, ZHENGZHOU UNIVERSITY, Henan 450003, China)
[4](STATE KEY LAB OF PROCESSORS, Institute of Computing Technology, Chinese Academy of Sciences, Beijing 100190, China)
[5](University of Chinese Academy of Sciences, Beijing 100049, China)
Corresponding author:Tang Dan, E-mail: tangdan@bosc.ac.cn

**Abstract**:　Considering the high-performance and low-power requirements of edge AI, this study designs a specialized instruction set processor for edge AI based on the RISC-V instruction set architecture, addressing practical issues in digital signal processing for edge devices. This design enhances the execution efficiency of edge AI and reduces its energy consumption with limited hardware overhead, meeting the demands for efficient large language model (LLM) inference computation in edge AI applications. The main contributions of this paper are as follows: For the characteristics of large language models, custom instructions were extended based on the RISC-V instruction set to perform vector dot product calculations, accelerating the computation of large language models on dedicated vector dot product acceleration hardware. Based on the open-source high-performance RISC-V processor core XiangShan Nanhu architecture, the vector dot product specialized instruction set processor Nanhu-vdot was implemented, which adds vector dot product calculation units and pipeline processing logic on top of the XiangShan Nanhu.The Nanhu-vdot underwent FPGA hardware testing, achieving over four times the speed of scalar methods in vector dot product computation. Using a hardware-software co-design approach for second-generation Generative Pre-Trained Transformer (GPT-2) model inference, the speed improved by approximately 30% compared to pure software implementation with almost no additional consumption of hardware resources and power consumption.

---

随着人工智能的迅速发展,全球对计算能力的需求呈现出爆炸式增长。这种趋势不仅推动了核心计算资源的升级,也引发了对更高效、更低延迟的计算模式的探索。边缘 AI(AI Edge)是人工智能 (AI) 与边缘计算交叉的先进技术,这一概念源于 AI 从云端向边缘下沉的分布式计算范式转变。在边缘计算架构中,数据不再需要全部上传至远程云服务器进行处理,而是能够在数据产生的本地环境中进行实时分析与决策。这种方式不仅提高了数据处理的效率,还减少了网络带宽的消耗,并增强了系统的可靠性和安全性。这在智能监控、自动驾驶、实时医疗诊断或工业自动化控制等应用场景中尤其重要[1]。

端侧大模型是部署在边缘设备(如智能手机、物联网设备、嵌入式系统等)上的大型机器学习模型,通过模型压缩、剪枝、量化等优化技术,使其能够在计算和存储资源有限的设备上高效运行[2,3]。与云端推理相比,端侧大模型提供了实时低延迟的响应能力,同时增强了数据隐私保护和离线处理的能力,适用于语音识别、图像处理、增强现实、智能家居等需要快速响应和高隐私的应用场景[4]。边缘计算技术在这一过程中起到了关键作用,成为释放端侧大模型潜力的核心驱动力。通过边缘计算,设备能够更好地管理和分配资源,进一步提升端侧大模型的性能和应用广度,提升了用户体验和系统的自主性[5]。

指令集扩展(ISA 扩展)是指在现有处理器的指令集架构(ISA)基础上增加新的指令,以提升特定任务的性能或支持新的功能。指令集是处理器能够理解和执行的命令集合,定义了处理器与软件之间的接口[6]。通过扩展指令集,可以为特定的应用场景或算法提供优化支持,从而提高计算效率或减少代码复杂性[7]。例如 Intel SSE(Streaming SIMD Extensions)是 Intel 在其 x86 处理器上进行的多次指令集扩展,增加了单指令多数据(SIMD)指令[8],极大提升了多媒体、图形处理和科学计算的性能。

矢量点积(Vector Dot Product)在大语言模型推理中是不可或缺的,它在神经网络的基本计算、注意力机制的实现、向量相似性计算以及高效并行化方面都起着至关重要的作用[9]。点积运算的性能对整个模型推理的效率和效果都有直接的影响,因此在大语言模型推理中占据了非常重要的地位。

传统的纯 Python 语言或 C/C++语言实现点积运算需要大量的数据访存操作,访存指令多、执行时间长、实时性较差,不能满足大模型推理的计算需求。而基于 GPU、TPU[10]的大模型推理加速,虽然解决了实时性问题,但是存在部署代价大、能效比低的缺陷[11]。因此,设计一种既能加速运算,高效且具备软件灵活性、利于部署应用的硬件架构是有意义的。

RISC-V 是一种基于精简指令集计算(Reduced Instruction Set Computing, RISC)原则的开源指令集架构[12]。RISC-V 的出现不仅避免了 X86 和 ARM 为了向后兼容而愈发臃肿的诟病[13],而且 RISC-V 处理器具有低成本、可灵活扩展等特性[14],在边缘计算领域能够较好地进行性能与功耗平衡[15]。在芯片国产化的趋势下,RISC-V 指令集由于其开源的特性,具有巨大的研究价值和广阔的应用前景[16]。

针对上述问题,本文基于专用指令集加速器的技术路线,提出一种基于 RISC-V 的矢量点积计算加速单元。目的是希望在提高端侧大模型推理算法硬件执行效率的基础上,减少额外的硬件资源和功耗开销。主要工作包括以下几个方面:

1) 针对大语言模型的特点,基于 RISC-V 指令集扩展了自定义指令,实现矢量点积计算,并在专用的矢量点积加速硬件上加速大语言模型的运算;

2) 基于开源的高性能 RISC-V 处理器核高性能处理器"香山"(nanhu 版本)[17],开发了包含矢量点积专用指令集的处理器 nanhu-vdot,在高性能处理器"香山"(nanhu 版本)的基础上增加了矢量点积计算单元和流水线处理逻辑;

3) 对 nanhu-vdot 进行了 FPGA 硬件测试,结果显示 nanhu-vdot 对比"香山"(nanhu 版本)几乎没有增加额外的硬件资源和功耗消耗,nanhu-vdot 矢量点积运算速度相比标量方法提高了 4 倍以上,并且在使用软硬件协同方案进行第二代生成式预训练模型(GPT-2)[18]推理时,速度比纯软件实现提高了 30%。

本文第 1 节介绍 RISC-V 指令集扩展加速推理相关工作.第 2 节介本文所需的基础知识,包括大语言模型的机制和方法.第 3 节介绍本文构建的矢量点积专用指令集处理器的加速功能模块设计.第 4 节介绍本文采取的软硬件协同实现.第 5 借通过对比实验验证了本文的有效性.最后总结全文.

# 1 RISC-V 指令集扩展加速推理相关工作

RISC-V 的向量扩展 (RISC-V Vector Extension, RVV) [19]是目前最重要的用于加速 AI 和大模型推理的扩展之一。该扩展使得处理器能够高效处理矩阵运算和并行计算任务,这对于大规模神经网络推理至关重要。研究表明,通过使用 RVV,RISC-V 处理器在执行卷积神经网络(CNN)和变换模型(Transformer)的推理任务时,可以显著提升性能。为了支持广泛的操作,RVV 可能需要更多的硬件资源(如寄存器文件和控制逻辑),这可能导致更高的功耗和芯片面积消耗。RVV 的复杂性在于它需要处理器的各个部分进行协调,以支持灵活的向量长度和类型。这增加了硬件设计的复杂性,特别是在寄存器文件、内存接口和调度器的设计上。

除了通用的向量扩展外,还有一些专用加速器(Domain-Specific Accelerators, DSAs)被集成到 RISC-V 架构中,以加速特定的 AI 任务。这些加速器通常通过自定义指令集扩展 (Custom ISA Extensions) 来优化特定的算法,如卷积运算或注意力机制。专用加速器往往需要专门的硬件接口,Gemmini [20]是一个针对 RISC-V 的深度神经网络 (DNN) 加速器,它需要处理器核拥有 RoCC[21]加速器扩展接口才能使用,这增加了系统设计和验证的复杂性,特别是在需要与处理器核心进行高效通信时。此外,由于加速器和核心之间的带宽和延迟要求严格,这可能进一步增加设计挑战。

# 2 第二代生成式预训练(GPT-2)大语言模型算法分析

GPT-2 是 OpenAI 发布的一个自然语言生成模型,全称为"Generative Pretrained Transformer 2"。它是基于 Transformers 架构的第二代语言模型,旨在生成高质量的文本。GPT-2 以无监督学习的方式在大量互联网文本数据上进行预训练,并通过转移学习应用于各种下游任务,如文本生成、翻译、问答等。

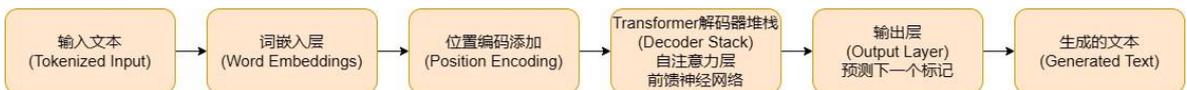

图 1 GPT-2 简化架构图

图 1 简单描述了 GPT-2 的工作原理,原始文本被分割成若干标记(tokens),每个标记代表一个词或子词;词嵌入层中每标记通过词嵌入层被转换为一个高维度向量,这些向量表示文本中的词在模型理解中的位置;由于 Transformers 缺乏对序列顺序的内在理解,位置编码被添加到每个词嵌入中,提供位置信息;多层的自注意力机制和前馈神经网络组成了模型的主要部分,每层通过自注意力机制(Self-Attention)处理输入序列,并生成中间表示,残差连接(Residual Connection)和层归一化(Layer Normalization)确保模型训练稳定性和性能;最后的输出是预测的下一个标记,模型根据这些预测逐步生成完整的文本。

## 2.1 神经网络全连接层

神经网络的全连接层(也称为密集层,Dense Layer)是一个非常重要的组件,在此层中矢量点积起到了关键作用[22,23]。全连接层的目的是对输入进行线性变换,然后应用激活函数。

它的基本计算公式为:

$$y = Wx + b \tag{1}$$

其中:$x$是输入向量,$W$是权重矩阵,$b$是偏置向量,$y$是输出向量。输入向量 $x$与权重矩阵$W$的乘法本质上是多个矢量点积的组合。权重矩阵$W$的每一行是一个权重向量$w_i$,与输入向量$x$进行点积,得到一个输出值$y_i$:

$$y_i = w_i \cdot x = \sum_{j=1}^{n} w_{ij} \cdot x_j \tag{2}$$

这个操作会为矩阵$W$的每一行计算一个点积,并将结果存储在输出向量$y$的相应位置。

## 2.2 注意力机制

在 Transformer 模型[24](如 GPT、BERT)中,注意力机制是一种关键组件[25]。注意力机制计算"注意力权重"(attention weights)允许模型在生成每个词语的过程中对输入序列的不同部分分配不同的权重,过程依

赖于查询向量（query）和键向量（key）之间的点积。这个点积计算表示输入序列中各个元素之间的相关性。通过这种方式，模型可以"注意"到序列中对当前任务最相关的部分。

$$Attention(Q, K, V) = softmax\left(\frac{Q K^T}{\sqrt{d_k}}\right) \cdot V \tag{3}$$

其中：$Q$是查询矩阵，$K$是键矩阵，$V$是值矩阵，$\sqrt{d_k}$是维度缩放的因子。注意力机制的公式中包含了矢量点积的计算，矢量点积 $Q K^T$ 的结果体现了查询与键的关联度，而这对于生成下一个词语时的决策非常关键[26]。

### 2.3 相似性计算

余弦相似度是矢量点积在相似性计算中最常见的应用之一[27]。当两个向量被归一化后，它们的点积值直接反映了它们之间的余弦相似度，这在推荐系统、信息检索和自然语言处理中的文本匹配等任务中非常重要。

### 2.4 优化模型推理性能

在大语言模型的推理过程中，许多计算瓶颈都是矩阵运算，点积是这些矩阵运算的核心。因此，点积计算的效率直接影响模型推理的性能。硬件加速器通常都专门优化了点积运算，提升了大语言模型的推理速度。

由上述可得矢量点积是大语言模型推理中的基本运算结构。

## 3 矢量点积专用指令集处理器的加速功能模块设计

本文设计使用高性能处理器"香山"(nanhu版本) SoC 作为硬件平台。在实现方式上，专用指令集处理器通常以协处理器的方式外挂在处理器上[28]，nanhu-vdot 没有采用这种方式，而是将矢量点积扩展指令与高性能处理器"香山"(nanhu版本)的流水线紧密、耦合。这样能够充分利用"香山"的现有译码逻辑、寄存器堆和功能单元，尽可能减少额外的面积开销。同时，也没有修改对外访问内存的总线宽度，目的是充分利用原有的访存模块逻辑，减少访存的硬件资源和功耗开销。

### 3.1 高性能处理器"香山"(nanhu版本) SoC平台

"香山"是一个基于 RISC-V RV64 开放指令集的可配置处理器核，使用 Chisel 语言[29]进行模块化设计，目前已经开源，主要用于学术研究。高性能处理器"香山"(nanhu 版本)采用了多级流水线设计，存储系统包括指令缓存、数据缓存和可配置的二级数据缓存，支持硬件数据预取[30]，并通过 AXI 总线与外部设备进行通信。高性能处理器"香山"(nanhu 版本)支持 RV64GCBK 扩展，支持虚实地址转换，包含页表缓冲以加速地址转换过程，支持 Sv39 分页方案。该处理器设计包括三个部分：前端（Frontend）负责分支预测和指令取指[31]，并将指令放置在译码缓冲区中；后端（Backend）负责执行指令，并从译码缓冲区中读出指令；访存单元（LSU）作为一个功能单元包含在后端流水线中，控制逻辑分布在流水线的各个部分。

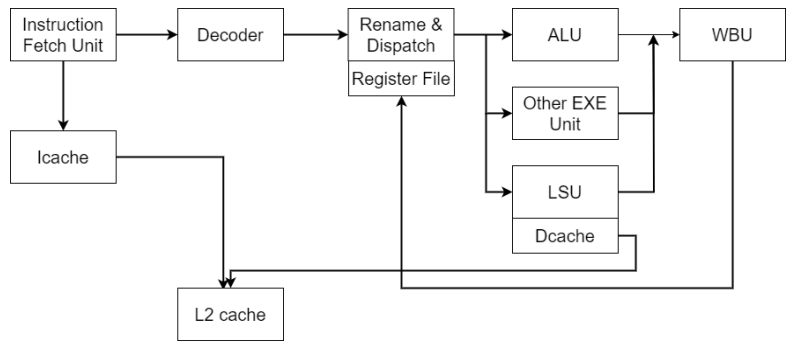

图 2 高性能处理器"香山"(nanhu 版本)结构

### 3.2 VDOTU执行模块

VDOTU 作为一个独立的模块位于流水线的执行级，与 ALU，LSU 等其他执行模块并列在执行级中进行数据的处理。通过在译码级生成的控制信号进行选择是否将操作数和操作码传递至 VDOTU 执行。VDOTU 是扩展指令的核心执行单元，采用 SIMD 向量化的执行方式。VDOTU 默认配置为 8bit 的整形计算，VDOTU

包含八路 8-bit 乘法器和七个加法器。输出采用 64-bit，与处理器的通用寄存器大小一致。

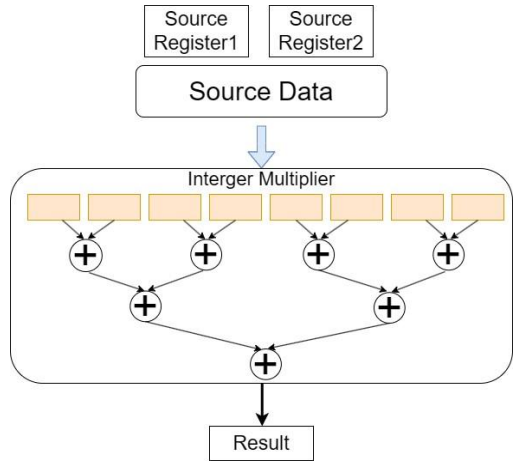

图 3 VDOTU 执行模块结构

## 4 软硬件协同实现

### 4.1 用V点积扩展指令替换GPT-2中V点积实现

当前大语言模型推理算法主要以纯软件或 GPU 实现，其中纯软件实现是指只利用处理器核中含有的取指、译码、执行、写回部件，运行 C 或 Python 代码，访存指令多、执行时间长、实时性较差；基于 GPU 实现，部署代价大、能效比低。本文设计在通用 CPU 系统中额外增加一个矢量点积计算执行单元，其使用软硬件协同方式，在执行单元模块中实现自定义指令的执行，从而实现大语言模型推理算法硬件上的加速。

硬件方面：编写矢量点积计算定制自定义扩展指令的单元设计代码，对矢量点积进行加速，与高性能处理器"香山"(nanhu 版本)一起编译，生成可仿真的比特流。

软件方面，修改 GPT-2 开源 C/C++代码，其中对于 int8 类型矢量点积计算部分通过汇编指令调用硬件执行单元,在调用硬件前后进行数据类型转换,最终通过硬件的加速计算得到文本输出。

### 4.2 自定义矢量点积指令扩展指令

在 RISC-V 指令集架构中，指令的编码格式设计需要兼顾指令的功能、硬件实现的复杂度和指令集的可扩展性。在进行大语言模型推理时，标准的 RISC-V 指令并不能完全满足算法向量化的要求。为了实现标准 RISC-V 不支持的功能和对大语言模型推理算法进行加速，本文新增了自定义矢量点积计算指令，并通过修改编译器、设计硬件电路的译码和执行模块实现了该指令。

矢量点积操作采用 RISC-V 指令的 R-type 译码模式，Inst[11:7]表示交换后数据写回的目的寄存器号，Inst[19:15]和 Inst[24:20]表示源操作数寄存器号，用于加载输入数据，共有两个输入寄存器号，即有两组输入数据。编码的前 7 位和后 7 位是固定的，这样可以确定该指令属于某个特定的指令类别或扩展指令集。具体来说，0001011 的后缀在 RISC-V 中通常用于自定义指令（custom-0 指令）。这种固定的编码模式可以有效区分不同类型的指令，并且可以保证自定义指令不会与标准指令集发生冲突。

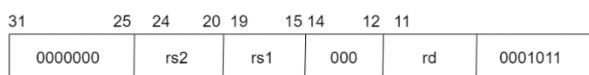

图 4 矢量点积计算指令编码格式

### 4.3 软件实现

本文设计优化 GPT-2 大模型推理中元素类型为 int8 的矢量点积计算实现。软硬件协同的矢量点积实现如算法 1 所示。

**算法 1** 软硬件协同的矢量点积实现

| | |
|---|---|
| 输入： | 两个包含 32 个 int8 类型元素的数据块（数组）X 和 Y |
| 输出： | 矢量点积的结果 |
| 步骤 1： | 按顺序把 X 数据块中的数每 32 个数存入 4 个通用寄存器 $x_i$，每个寄存器存 8 个 int8 数据 |
| 步骤 2： | 按顺序把 Y 数据块中的数每 32 个数存入 4 个通用寄存器 $y_i$，每个寄存器存 8 个 int8 数据 |
| 步骤 3： | 通用寄存器 $x_i$ 与对应的寄存器 $y_i$ 作为矢量点积计算源寄存器，通过混合汇编调用自定义指令，由扩展加速单元计算完成 4 个点积结果 |
| 步骤 4： | 由软件执行 4 个点积结果累加，并转换数据类型 |

## 5 FPGA 仿真及性能分析现

本文基于"香山"开源处理器核（nanhu 版本）架构的相关生态来构建实验环境。考虑到降低 FPGA 资源占用和仿真工程运行耗时，本文选用高性能处理器"香山"(nanhu 版本)的最小配置进行仿真，最小配置相比标准配置减少了译码单元数量、后端执行单元数量、去除了 L2 缓存。

### 5.1 实验配置

本文使用上海思尔芯技术股份有限公司 (S2C) 的 Prodigy™ S7-19PS Logic System[32]机箱作为实验平台进行 nanhu-vdot 的测试，该系统基于 Xilinx VU19P FPGA 开发板搭建。软件仿真环境为 Vivado 2020.2

### 5.2 高性能处理器"香山"(nanhu版本)仿真工程的构建

高性能处理器"香山"nanhu-vdot 的 FPGA 硬件测试，首先进行硬件测试环境的构建，主要包括：SoC 工程的构建和 FPGA 测试用例的生成。然后进行 nanhu-vdot 的执行效率和功耗的测试，对比结果分析。

高性能处理器"香山"nanhu-vdot 使用 Chisel HDL 语言进行开发，仿真时首先需要使用 Scala[33]的编译工具 Mill[34]将 Chisel 转换成 Firrtl 代码[35]然后再利用 Firrtl 编译器转换成 Verilog 顶层文件。高性能处理器"香山"nanhu-vdot 作为处理器无法独立地运行测试样例，它需要时钟，内存，外围接口电路等的配合才能作为一个完整的片上系统（System on Chip，SoC）执行任务。使用 Vivado 2020.2 构建高性能处理器"香山"nanhu-vdot SoC 工程，综合和实现后生成比特流文件，从而可以将高性能处理器"香山"nanhu-vdot 系统下载到 FPGA 开发板上。

### 5.3 FPGA 测试用例的生成

本文分别测试了高性能处理器"香山"nanhu-vdot 对比高性能处理器"香山"(nanhu 版本)计算矢量点积的速度和执行 GPT-2 三种大小的模型推理的速度。其中 GPT-2 的小型模型参数量为 117M、中型模型的参数量为 345M、大型模型的参数量为 774M，模型文件中参数的格式与本文硬件单元中的元素位宽一致。

表 1 GPT-2 测试模型参数

| Model | 参数量 | n_vocab | n_ctx | n_embd | n_head | n_layer | qntvr |
|---|---|---|---|---|---|---|---|
| Small | 117M | 50257 | 1024 | 768 | 12 | 12 | 2 |
| Medium | 345M | 50257 | 1024 | 1024 | 16 | 24 | 2 |
| Large | 774M | 50257 | 1024 | 1280 | 20 | 36 | 2 |

其中 n_vocab 表示词汇表的大小，也就是模型能够识别的标记（tokens）的数量；n_ctx 表示上下文窗口

大小，即模型可以考虑的最大序列长度（标记数）；n_embd 表示嵌入维度的大小，即每个标记被嵌入为一个向量时的维度数；n_head 表示注意力头的数量；n_layer 表示 Transformer 解码器的层数；qntvr 表示注意力矩阵中的因子数。

将 GPT-2 模型推理文件和模型文件处理为操作系统可执行文件，并通过测试接口外设将操作系统可执行文件传输至硬件电路的内存，后续可以在硬件电路的内存中执行操作系统可执行文件，从而启动操作系统，在操作系统中执行模型推理程序，实现在电路设计的测试过程中执行模型推理的目的。

利用大模型推理辅助 nanhu-vdot 测试的过程中，模型文件和模型推理程序需运行在文件系统中。临时文件系统包括了操作系统必要的驱动程序、工具和配置文件，本文将推理程序和模型文件导入 Linux 临时文件系统，得到操作系统可执行文件；。通过 Jtag 接口外设将操作系统可执行文件传输至 S7-19PS 的内存，以便后续在内存中启动操作系统并执行模型推理过程。

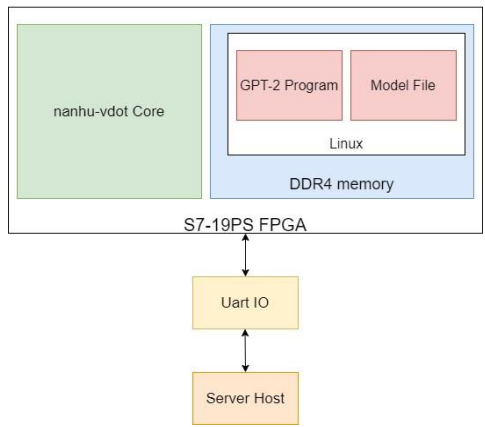

图 5 FPGA GPT-2 推理测试系统结构

**5.4 系统硬件仿真与结果分析**

**5.4.1 硬件资源消耗对比**

基于硬件实验环境，在 Vivado 工具上将高性能处理器"香山"nanhu-vdot 和高性能处理器"香山"(nanhu 版本)最小配置进行综合和实现，将实现的数据作为对比，硬件资源消耗如表 1 所示：

表 2 "香山"(nanhu 版本)与 nanhu-vdot 硬件资源消耗对比

|  | LUT | Flip Flop | BRAMs |
|---|---|---|---|
| nanhu | 564211 | 278746 | 436.5 |
| nanhu-vdot | 579888 | 281232 | 436.5 |

nanhu-vdot 相比"香山"(nanhu 版本)增加 15677 个 LUT 单元，占比 2.8%，增加 2486 个 Flip-Flop 单元，占比 0.9%，BRAMs 未增加。由此可见本设计相比于"香山"(nanhu 版本)硬件资源的开销没有大幅增加。

"香山"(nanhu 版本)的功耗开销为 8.454W，nanhu-vdot 功耗为 8.494W。nanhu-vdot 相比于"香山"(nanhu 版本)的功耗仅增加 0.5%，几乎不会增加处理器的功耗。

**5.4.2 矢量点积计算速度对比**

矢量点积执行效率方面，nanhu-vdot 与"香山"(nanhu 版本)执行 50000 次该计算的耗时分别为 99.96 和 24.72ms，可见矢量点积扩展对该计算指令具有 4 倍以上的加速比，对比标量方法计算速度显著提升。

### 5.4.3 GPT-2 推理速度对比

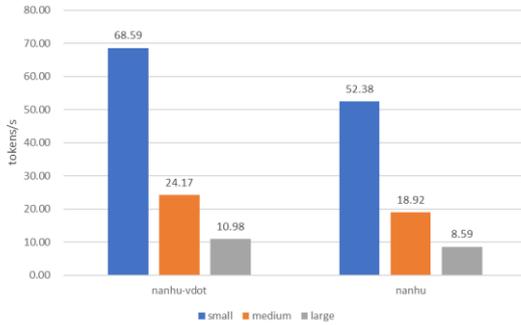 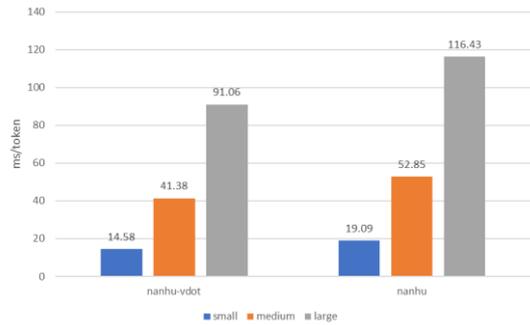

(a)GPT-2 推理速度结果　　　　　　　　(b) GPT-2 推理耗时结果

图 6 nanhu-vdot 对比"香山"（nanhu 版本)计算 GPT-2 大语言模型推理速度效率对比

大模型执行效率方面，对 GPT-2 小型模型、中型模型、大型模型的推理速度提升分别为 30.9%、27.8%、27.9%。因此，本文提出的设计可以提升处理器核对于大模型推理的处理效率，这对边缘计算设备的性能提升十分重要。

## 6 总　结

针对边缘 AI 的自主性和实时性的需求，本文设计并实现了一种针对边缘 AI 应用的专用指令集处理器。通过扩展 RISC-V 指令集，增加自定义矢量点积指令，本文提出的设计在高性能处理器"香山"（nanhu 版本）架构基础上实现了 nanhu-vdot 处理器，旨在满足高效大语言模型（LLM）推理计算的需求。该处理器集成了矢量点积计算单元和流水线处理逻辑，大幅提升了计算效率。在 FPGA 硬件测试中，矢量点积运算速度较标量方法提高了 4 倍以上，使用软硬件协同方案对 GPT-2 模型进行推理时，速度相比纯软件实现提高了约 30%。这一设计在不显著增加硬件资源和能耗的情况下，显著提升了边缘 AI 设备的大语言模型推理效率，满足了边缘 AI 对高性能与低功耗的要求。

**References**:
[1] Li Y, Zhu J, Fu Y, et al. Circular Reconfigurable Parallel Processor for Edge Computing: Industrial Product∗[C]//2024 ACM/IEEE 51st Annual International Symposium on Computer Architecture (ISCA). IEEE, 2024: 863-875.
[2] Daghero F, Pagliari D J, Poncino M. Energy-efficient deep learning inference on edge devices[M]//Advances in Computers. Elsevier, 2021, 122: 247-301
[3] Dhar S, Guo J, Liu J, et al. A survey of on-device machine learning: An algorithms and learning theory perspective[J]. ACM Transactions on Internet of Things, 2021, 2(3): 1-49.
[4] V. N. Chander and K. Varghese, "A Soft RISC-V Vector Processor for Edge-AI," 2022 35th International Conference on VLSI Design and 2022 21st International Conference on Embedded Systems (VLSID), Bangalore，India, 2022, pp. 263-268, doi: 10.1109/VLSID2022.2022.00058．
[5] Singh R, Gill S S. Edge AI: a survey[J]. Internet of Things and Cyber-Physical Systems, 2023, 3: 71-92.
[6] T. He, X. Chen and G. Wang, "Research on Open Source Processor and Analysis of Current Development Dilemma Based on RISC-V," 2023 8th International Conference on Computer and Communication Systems (ICCCS), Guangzhou, China, 2023, pp. 768-774, doi: 10.1109/ICCCS57501.2023.10151171．
[7] Yue Gao, Wei Qian, and Enfang Cui. 2023. RISC-V ISA Extension Toolchain Supports: A Survey. In Proceedings of the 2023 4th International Conference on Computing, Networks and Internet of Things (CNIOT '23)．
[8] Kusswurm D, Kusswurm D. Streaming simd extensions[J]. Modern X86 Assembly Language Programming: 32-bit, 64-bit, SSE, and AVX, 2014: 179-206.
[9] Robbert Emery. " How AI and ML Applications Will Benefit from Vector Processing." enterpriseai.news, 31 July. 2020, https://www.enterpriseai.news/2020/07/31/how-ai-and-ml-applications-will-benefit-from-vector-processing.

[10] Jouppi N P, Young C, Patil N, et al. In-datacenter performance analysis of a tensor processing unit[C]//Proceedings of the 44th annual international symposium on computer architecture. 2017: 1-12.
[11] Zhou, L., Zhao, Z. Q., Pan, G. T., et al. Design of a Graph Convolutional Neural Network Accelerator Based on RISC-V [J]. Computer Engineering and Science, 2023, 45(12): 2113-2120.
[12] Liu, C., Wu, Y. J., Wu, J. Z., Zhao, C. A Review of RISC-V Instruction Set Architecture Research [J]. Journal of Software, 2021, 32(12): 3992-4024.
[13] Li, F., Guo, S. Z., Hao, J. W., et al. Implementation of a Basic Mathematical Library for RISC-V [J]. Journal of Electronics, 2024, 52(05): 1633-1647.
[14] E. Cui, T. Li and Q. Wei, "RISC-V Instruction Set Architecture Extensions: A Survey," in IEEE Access, vol. 11, pp. 24696-24711, 2023, doi: 10.1109/ACCESS.2023.3246491
[15] E. Torres-Sánchez, J. Alastruey-Benedé and E. Torres-Moreno, "Developing an AI IoT application with open software on a RISC-V SoC," 2020 XXXV Conference on Design of Circuits and Integrated Systems (DCIS), Segovia, Spain, 2020, pp. 1-6, doi: 10.1109/DCIS51330.2020.9268645.
[16] Valerii Haidarzhy. "RISC-V Unleashed: The Definitive Guide to Next-Gen Computing." Sirin Software, 16 Apr. 2024, https://sirinsoftware.com/blog/risc-v-unleashed-the-definitive-guide-to-next-gen-computing.
[17] Y. Xu *et al*., "Towards Developing High Performance RISC-V Processors Using Agile Methodology," *2022 55th IEEE/ACM International Symposium on Microarchitecture (MICRO)*, Chicago, IL, USA, 2022, pp. 1178-1199, doi: 10.1109/MICRO56248.2022.00080.
[18] Radford A, Wu J, Child R, et al. Language models are unsupervised multitask learners[J]. OpenAI blog, 2019, 1(8): 9.
[19] "Working draft of the proposed RISC-V V vector extension", [online] Available: https://github.com/riscv/riscv-v-spec.
[20] Genc H, Kim S, Amid A, et al. Gemmini: Enabling systematic deep-learning architecture evaluation via full-stack integration[C]//2021 58th ACM/IEEE Design Automation Conference (DAC). IEEE, 2021: 769-774
[21] Zhao J, Korpan B, Gonzalez A, et al. Sonicboom: The 3rd generation berkeley out-of-order machine[C]//Fourth Workshop on Computer Architecture Research with RISC-V. 2020, 5: 1-7.
[22] Basha S H S, Dubey S R, Pulabaigari V, et al. Impact of fully connected layers on performance of convolutional neural networks for image classification[J]. Neurocomputing, 2020, 378: 112-119.
[23] Shalev-Shwartz S, Ben-David S. Understanding machine learning: From theory to algorithms[M]. Cambridge university press, 2014.
[24] D'Souza, Jennifer. (2023). A Review of Transformer Models. 10.48366/r640001.
[25] Vaswani A, Shazeer N, Parmar N, et al. Attention Is All You Need.(Nips), 2017[J]. arXiv preprint arXiv:1706.03762, 2017, 10: S0140525X16001837.
[26] DeRose J F, Wang J, Berger M. Attention flows: Analyzing and comparing attention mechanisms in language models[J]. IEEE Transactions on Visualization and Computer Graphics, 2020, 27(2): 1160-1170.
[27] Uprety S, Jaiswal A K, Liu H, et al. Investigating Context Effects in Similarity Judgements in Large Language Models[J]. arXiv preprint arXiv:2408.10711, 2024.
[28] Jiang, S. J. Design of an FFT-Specific Instruction Set Processor Based on RISC-V [D]. South China University of Technology, 2023. DOI: 10.27151/d.cnki.ghnlu.2023.005128.
[29] Bachrach J, Vo H, Richards B, et al. Chisel: constructing hardware in a scala embedded language[C]//Proceedings of the 49th Annual Design Automation Conference. 2012: 1216-1225.
[30] Y. Zhu, J. Zheng, S. Ding and L. Li, "Hardware Data Prefetch for XiangShan Processor," 2022 7th International Conference on Integrated Circuits and Microsystems (ICICM), Xi'an, China, 2022, pp. 394-397, doi: 10.1109/ICICM56102.2022.10011259.
[31] Jiangrui Zou, Dan Tang, Ye Cai, Zusong Li, "A design of fetch target buffer implemented on XiangShan processor," Proc. SPIE 12303, International Conference on Cloud Computing, Internet of Things, and Computer Applications (CICA 2022), 123032S (28 July 2022);
[32] Xilinx, "Product Overview: 1-1dt42z7 Development Board," accessed August 25, 2024. [Online]. Available: https://china.xilinx.com/products/boards-and-kits/1-1dt42z7.html.
[33] Li P S, Izraelevitz A M, Bachrach J. Specification for the FIRRTL Language[J]. EECS Department, University of California, Berkeley, Tech. Rep. UCB/EECS-2016-9, 2016.
[34] "Java Introduction to Mill." *Mill Documentation*, Mill Build, Version 0.11.12, https://mill-build.org/mill/0.11.12/Java_Intro_to_Mill.html. Accessed 25 August 2024.

[35] A. Izraelevitz *et al*., "Reusability is FIRRTL ground: Hardware construction languages, compiler frameworks, and transformations," 2017 IEEE/ACM International Conference on Computer-Aided Design (ICCAD), Irvine, CA, USA, 2017, pp. 209-216, doi: 10.1109/ICCAD.2017.8203780.

**附中文参考文献:**

[11] 周理,赵祉乔,潘国腾,等.基于 RISC-V 的图卷积神经网络加速器设计[J].计算机工程与科学,2023,45(12):2113-2120.
[12] 刘畅,武延军,吴敬征,赵琛. RISC-V 指令集架构研究综述.软件学报,2021,32(12):3992-4024.
[13] 李飞,郭绍忠,郝江伟,等.面向 RISC-V 的基础数学库实现[J].电子学报,2024,52(05):1633-1647.
[28] 江世杰.基于 RISC-V的 FFT 专用指令集处理器设计[D].华南理工大学,2023.DOI: 10.27151/d.cnki.ghnlu.2023.005128